\documentclass[10pt, twocolumn, dvipsnames]{article}
\usepackage{cite, here}
\usepackage{algorithm}
\usepackage{algpseudocode}
\usepackage{amsmath,amssymb,amsfonts,xcolor}
\usepackage{amsmath,amssymb}
\usepackage[utf8]{inputenc}

\usepackage{bm}

\newtheorem{remark}{\sc Remark}
\newtheorem{proposition}{\sc Proposition}
\newtheorem{definition}{\sc Definition}
\usepackage{tikz}
\usetikzlibrary{arrows,shapes,calc}
\usepackage{pgfplots}
\usetikzlibrary{positioning} 

\tikzstyle{solid}=                   [dash pattern=]
\tikzstyle{dotted}=                  [dash pattern=on \pgflinewidth off 2pt]
\tikzstyle{densely dotted}=          [dash pattern=on \pgflinewidth off 1pt]
\tikzstyle{loosely dotted}=          [dash pattern=on \pgflinewidth off 4pt]
\tikzstyle{dashed}=                  [dash pattern=on 3pt off 3pt]
\tikzstyle{densely dashed}=          [dash pattern=on 3pt off 2pt]
\tikzstyle{loosely dashed}=          [dash pattern=on 3pt off 6pt]
\tikzstyle{dashdotted}=              [dash pattern=on 3pt off 2pt on \the\pgflinewidth off 2pt]
\tikzstyle{densely dashdotted}=      [dash pattern=on 3pt off 1pt on \the\pgflinewidth off 1pt]
\tikzstyle{loosely dashdotted}=      [dash pattern=on 3pt off 4pt on \the\pgflinewidth off 4pt]

\usepackage{graphicx}      

\usepackage{graphicx}
\usepackage{textcomp}
\def\BibTeX{{\rm B\kern-.05em{\sc i\kern-.025em b}\kern-.08em
    T\kern-.1667em\lower.7ex\hbox{E}\kern-.125emX}}
\begin{document}
\title{Partial Extended Observability Certification and Optimal Design of  Moving Horizon Estimators Under Uncertainties}
\author{Mazen Alamir\thanks{Mazen Alamir is with CNRS/Gipsa-lab, Grenoble-INP, University of Grenoble Alpes, France. (e-mail: mazen.alamir@grenoble-inp.fr)}
\thanks{}}

\maketitle

\begin{abstract}
This paper addresses the observability analysis and the optimal design of observation parameters in the presence of noisy measurements and parametric uncertainties.  The main underlying frameworks are the nonlinear constrained moving horizon estimator design and the probabilistic certification via randomized optimization.  As the perfect observability concept is not relevant under the considered uncertain and noisy context, the notion of almost $\epsilon$-observability is introduced and a systematic procedure to assess its satisfaction for a given system with a priori known measurement noise statistics and parameter discrepancy is sketched. A nice feature in the proposed framework is that the observability is not necessarily defined as the ability to reconstruct the whole state, rather, the more general concept of  observation-target quantities is used so that one can analyze the precision with which specific chosen expressions of the state and the parameters can be reconstructed. The overall framework is exposed and validated through an illustrative example. 
\end{abstract}


\section{Introduction}
In nowadays data-focused period, there are still many engineering problems that require the design of state observers in the traditional sense \cite{Gildas2007}, namely those which are based on knowledge-driven models. Indeed, data-driven models only involve those variables for which sensors are available on-board. Engineering problems however,  involve quite often formulations that contain references to real-life variables that are not directly accessible via sensors. These physically meaningful non measured variables are used in the expression of some constraints and/or performance related computation and therefore need to be reconstructed  based on knowledge-based models.  
\\ \ \\ 
Knowledge-based models can be written elegantly using well motivated functional terms but they generally involve parameters with perfectly identified role but rather badly known values. The lack of precise knowledge of these values may have drastic consequences on the quality of the state observers that would be designed to dynamically inverse these models. The unavoidable measurement noise affecting the real-life sensors has comparable effects. \\ \ \\ 
When facing such situations, the available solutions can be split into the following {\color{black} three} categories: 
\\ \ \\ 
In the first category, only measurement noise with zero mean is considered. The observability is then analyzed in noise-free context using Lie-algebra local tools \cite{Hermann77} or thanks to structural dedicated high gain \cite{Astolfi15} or sliding modes \cite{Drakunov07} observability assumptions and the observer is then simulated in the presence of measurement noise in order to show that the dynamically estimated state chatters around its true trajectory. 
\\ \ \\ 
In the second category, parametric uncertainties are considered. Typical frameworks involve interval analysis \cite{MESLEM20089667,EFIMOV2013200} which is conducted to derive worst-case bounds on the estimation error \cite{ALAMO20051035} at the price of structural assumptions that generally involve the existence of bounding behavior for the uncertain system \cite{Li2012}. Such assumptions are rarely satisfied and when they are, they are limited to restricted observation intervals in order to avoid worst case error propagation that might quickly lead to useless bounds.  
\\ \ \\ 
{\color{black} The third category is related to moving horizon estimation schemes \cite{Muller2017automatica, Knufer2018CDC, Alessandri2008, Wynn2014, Michalska1995,Muske1995,Rao2000,Alamir2007obs}. This approach corresponds to a very rich literature starting from the early works \cite{Michalska1995, Muske1995} were the foundations of this optimization-based estimation approach have been laid in the deterministic setting. Since,  many improving refinements  have been proposed (see \cite{Rao2000} and the reference therein), including the proof of convergence of the deterministic settings in a real-time iteration framework \cite{Wynn2014} where only a single iteration of the optimization process is applied to update the estimates. \\ \ \\ As far as uncertainty handling is concerned, one of the more advanced MHE frameworks has been recently proposed in \cite{Muller2017automatica, Knufer2018CDC} which improved a family of  early works \cite{Alessandri2008, Alessandri2010, Rawlings2012, Ji2016} that mainly derive a gain-bounded error between the additive bounded disturbance term and the resulting estimation error. The results of  \cite{Muller2017automatica, Knufer2018CDC} extended the previous results to models with non necessarily additive disturbances term via less conservative estimation of the disturbance-to-error gain. A nice feature of the proposed approach is the guarantee that when the disturbance term vanishes and in the absence of noise, one gets an asymptotic convergence of the estimation error to $0$. \\ \ \\ 
The positioning of the present contribution compared to the previously cited MHE works can be summarized by the following items:
\begin{enumerate}
\item As far as the estimation error bound is concerned in MHE works, existing results are based on a worst-case analysis in which standard Lipschitz-like upper bounds are successively introduced in the derivation of the final bound. Such upper bounds, as acknowledged by the author of \cite{Muller2017automatica}, are too pessimistic if not useless in some cases. This might be the price to pay in order to derive convergence results in the case where the disturbances vanish. \\ \ \\ 
\textit{In the present contribution, a different angle of attack is chosen by focusing the  interest on the computation of tight probabilistic assessment of the estimation error excursion during the estimation process. This is done using a formulation that enables probabilistic certification tools to be deployed. Moreover, the paper is focused on parametric uncertainties rather than disturbance rejection making the concept of vanishing disturbances inappropriate as it is irrelevant to consider temporarily vanishing uncertainties}. \\
\item In all previous MHE works, the underlying assumption is that the system is globally detectable in the conventional sense. This means that the underlying paradigm is that of total state (or extended state) detectability. The analysis of partial reconstruction of specific functions of the problem's variables and parameters is not a straightforward extension and needs new results to be worked out from scratch. \\ \ \\
\textit{By opposition, the present contribution is totally built around the concept of partial estimation and the numerical investigation underlines its relevance by showinf that different precision might be achieved for different estimation targets.}
\end{enumerate}
}
\ \\
\noindent The first item in the above argumentation is one instance among many others of a new paradigm that is conveyed by the emergence of the probabilistic certification ideas and tools \cite{Calafiore2006,alamo2009,Calafiore2010}. This paradigm probably came as a reaction to the accumulation of  failures of {\em provable} solutions in addressing the real-world problems. The fact that the industrial world became more prone than before to welcome the advanced control solutions probably accelerated this shift if not made it unavoidable. In a nutshell, the probabilistic certification framework gives the minimal number of scenarios one has to randomly draw in order to check whether some statement can be certified with a given probability in an uncertain context. The higher the probability one is seeking to assess is, the larger  the number of scenarios to be tested with success should be. 
\\ \ \\ 
Generally speaking the probabilistic certification is a paradigm that can be invoked each time one would like to evaluate the veracity of a statement in a no more binary (True/False) way. The statement regarding observability makes no exception and this is precisely the starting point of the present contribution. Indeed, in the presence of noise and parameters mismatch, one should expect that for a any level of uncertainties and even in the absence  of measurement noise, some close states (lying in some $\epsilon$-neighborhood of the current state)  can become {\em indistinguishable from the only measurements}. Such an assertion itself can be stated in a probabilistic way. Moreover, the size of the $\epsilon$-neighborhood that is used in the formulation might be impacted by the choice of the observer's parameters.
\\ \ \\ 
It is worth underlying here that this paper is not about proposing a new numerical scheme to implement MHE. This can be efficiently done by many available tools \cite{Ferreau12,KUHL201171,Andersson2019}. {\color{black} The contribution aims at providing a method for the estimation of the size of possible regions of partial indistinguishability rather than a new implementation or a new cost formulation of MHEs. In other words, when the proposed approach fires a high risk of indistinguishability, this risk materializes {\bf regardless of the formulation} of the cost function that underlines the MHE. To this respect, the proposed methodology can be used in the top of any formulation as an outer hyper-parameter optimization step. }
\\ \ \\ 
Beside the formulation of the traditional observability analysis via randomized certification, this paper looks for a more general problem, namely the partial observability/identifiability problem. The question this paper answers in a probabilistic sense is the following: 
\\ \ \\ 
\begin{minipage}{0.02\textwidth}
\color{Brown}
\rule{1mm}{22mm}
\end{minipage}
\begin{minipage}{0.45\textwidth}
\color{Brown}
{\bf  Partial Extended Observability Problem}
{\em Given an expression $z=T(x,p)$ of the state $x$ and the parameter vector $p$, what is the precision to which $z$ can be dynamically reconstructed by using the available noisy measurements? }
\end{minipage}
\\ \ \\ \ \\
The word partial in the above definition refer to the fact that the map $T$ returns an expression that might contain only a part of the state, a part of the parameter vector or some combination of $x$ and $p$ that by no means enable to reconstruct $x$ or $p$ or both. A simple example that underlines the relevance of such a framework is the situation where one would like to reconstruct a feedback law $K(x,p)$ that is defined should the state and the parameters be known. In this case, one is not interested in reconstructing neither $x$ nor $p$ but only the expression $K(x,p)$. The cases where $z=x$ and $z=p$ recover the standard state and identification problem respectively. The case where $T(x,p)=p_i$ where $p_i$ is a specific component of $p$ corresponds to the situation where one of the parameters needs to be reconstructed for whatever reason while the reconstruction of the whole vector $p$ is not useful. To say it shortly one would be satisfied with the extended state/parameter vector being non precisely reconstructible from the measurement {\bf provided} that the expression of interest $z=T(x,u)$, called hereafter the {\em observation target}, can be reconstructed with acceptable precision. 
\\ \ \\ 
{\color{black} The idea of estimating a function of the system's variable and not the whole state is not new as it is in the heart of the investigations regarding the so-called functional observers (see \cite{Trinh2012, KRAVARIS2016505, TEH20131169} and the references therein). Although the majority of the related works concern linear systems or linear functions of the state to be reconstructed rather than the whole state, some few works apply to nonlinear dynamics. Almost invariably, the solutions amount at constructing a reduced order observer that is build after exhibiting a lower dimensional dynamics having the same dimension as  the function to be estimated. The settings are deterministic and the underlying construction heavily relies on structural assumptions on the mathematical equations of the model. Roughly speaking, the functional observer literature tries to do \textit{observer's business as usual}  although on reduced dimensional dynamics via structural \textit{matching conditions}. That is the reason why it has been chosen here to avoid the explicit use of the terminology of \textit{functional observer} in order to prevent misleading connections with such quite a different approach.}\\ \ \\ 
This paper is organized as follows: First some notation and definitions are introduced in Section \ref{sec-defnot} together with basic recalls regarding the definition of Moving Horizon Estimators (MHE). Section \ref{problemstatement} introduces the concept of $\epsilon$-observability of an observation target and shows that its satisfaction requires a robust constraints satisfaction condition. This condition is relaxed in Section \ref{secproba} which introduces the less stringent concept of almost $\epsilon$-observability and shows how this concept can be assessed using randomized optimization frameworks. Section \ref{secNI} proposes an illustrative example that precisely implements the framework on a simple example in order to show its effectiveness in investigating the degree of partial observability of a nonlinear system under measurement noise and parametric uncertainties. Finally Section \ref{secConc} concludes the paper and gives some suggestions for future investigation. 
\section{Definitions and Notation} \label{sec-defnot}
\noindent We consider nonlinear systems that are governed by a dynamical equation of the form:
\begin{equation}
x_{k+1}=f(x_k,u_k,p) \label{model}
\end{equation}
where $x_k\in \mathbb X\subset \mathbb{R}^{n_x}$, $u_k\in \mathbb U\subset \mathbb{R}^{n_u}$  stand for the values, at sampling instant $k$, of the state and the input  vectors respectively while $p\in \mathbb P$ stands for the vector of imperfectly known parameters of the model. It is also assumed that some vector of $n_y$ measurements (including the input) is available that is linked to the state and the control vector through noisy measurement, namely:
\begin{equation}
y_k = h(x_k,u_k,p)+\nu_k \label{defdeh}
\end{equation}
where $y_k\in \mathbb{R}^{n_y}$ stands for the sensor output at instant $k$ which is corrupted by the noise realization $\nu_k$. 
\\ \ \\ 
We assume that one is interested in reconstructing the value of some so-called {\bf  observation-target} variable\footnote{Note that using $T(x,p)=x$ [resp. $T(x,p)=(x,p)$] enables the common state observation and extended state/parameter observation to be recovered without loss of generality. }:
\begin{equation}
z = T(x,p) \label{defdezT}
\end{equation}
where $T: \mathbb R^{n_x}\times \mathbb R^{n_p}\rightarrow \mathbb{R}^{n_z}$ is some known map. 
\\ \ \\ 
Given any vector signal $$s = \begin{bmatrix}
 s^{[1]}\cr \vdots \cr  s^{[n_s]}
\end{bmatrix} \in \mathbb{R}^{n_s}$$ the forward/backward  profiles of $s$ at some instant $k$ over some window of length $N$ (in terms of sampling periods) is denoted as follows :
\begin{align}
\bm s^+_k&:= \begin{bmatrix}
s_k & s_{k+1} & \dots & s_{k+N-1} 
\end{bmatrix}\in \bigl[\mathbb{R}^{n_s}\bigr]^N \label{splus} \\
\bm s^-_k&:= \begin{bmatrix}
s_{k-N} & s_{k-N+1} & \dots & s_{k-1} 
\end{bmatrix}\in \bigl[\mathbb{R}^{n_s}\bigr]^N \label{smoins} 
\end{align}
which is a condensed expression gathering the profiles of all the components of $s$ that would be denoted individually according to:
\begin{align}
\bm s^{[i]+}_k&:= \begin{bmatrix}
s_k^{[i]} & s^{[i]}_{k+1} & \dots & s^{[i]}_{k+N-1} 
\end{bmatrix}\in \mathbb{R}^N \label{siplus} \\
\bm s^{[i]-}_k&:= \begin{bmatrix}
s_{k-N}^{[i]} & s^{[i]}_{k-N+1} & \dots & s^{[i]}_{k-1} 
\end{bmatrix}\in \mathbb{R}^N \label{simoins}
\end{align}
Using the above notation of control profile, it is now possible to define the future state prediction at instant $k+i$ given the current state $x_k$ at instant $k$, a given control profile $\bm u_k^+\in\mathbb U^N$ and some vector of parameters $p$ by using the following notation:
\begin{equation}
(\forall i\in \{0,\dots,N\})\qquad \hat x_{k+i} = X_i(x_k, \bm u_k^+, p) \label{eqflowX}
\end{equation}
{\color{black} which simply refers to the simulation based on (\ref{model}) with $x_k$ as initial state, $\bm u_k^+$ as control profile and $p$ as the parameter vector value.}
The same notation is used to refer to the predicted noise free output, namely:
\begin{equation}
(\forall i\in \{0,\dots,N\})\qquad \hat y_{k+i} = Y_i(x_k, \bm u_k^+, p) \label{eqflowY}
\end{equation}
Moreover, the resulting predicted output profile is simply denoted by:
\begin{align}
&\bm{Y}(x_k,\bm u^+_k,p):= \nonumber \\ &\begin{bmatrix}
 Y_0(x_k, \bm u_k^+, p)&\dots & Y_{N-1}(x_k, \bm u_k^+, p)
\end{bmatrix}  \in [\mathbb{R}^{n_y}]^N\label{bmyhat}
\end{align}
Given a candidate initial state $\xi$ at instant $k-N$, an input profile $\bm u_k^-$ that has been applied over the previous time interval $[k-N,k)$ and a candidate value $\hat p$ of parameter vector, a predicted trajectory  $\bm Y(\xi, \bm u_k^-, \hat p)$ can be obtained and compared to the truly measured one $\bm y^-_k$ to define the output prediction error profile by:
\begin{equation}
\bm e_k=\bm y_k^--\bm Y(\xi, \bm u_k^-, \hat p) \in \Bigl[\mathbb{R}^{n_y}\Bigr]^N
\end{equation}
\begin{figure}[H]
\begin{center}
\includegraphics[width=0.45\textwidth]{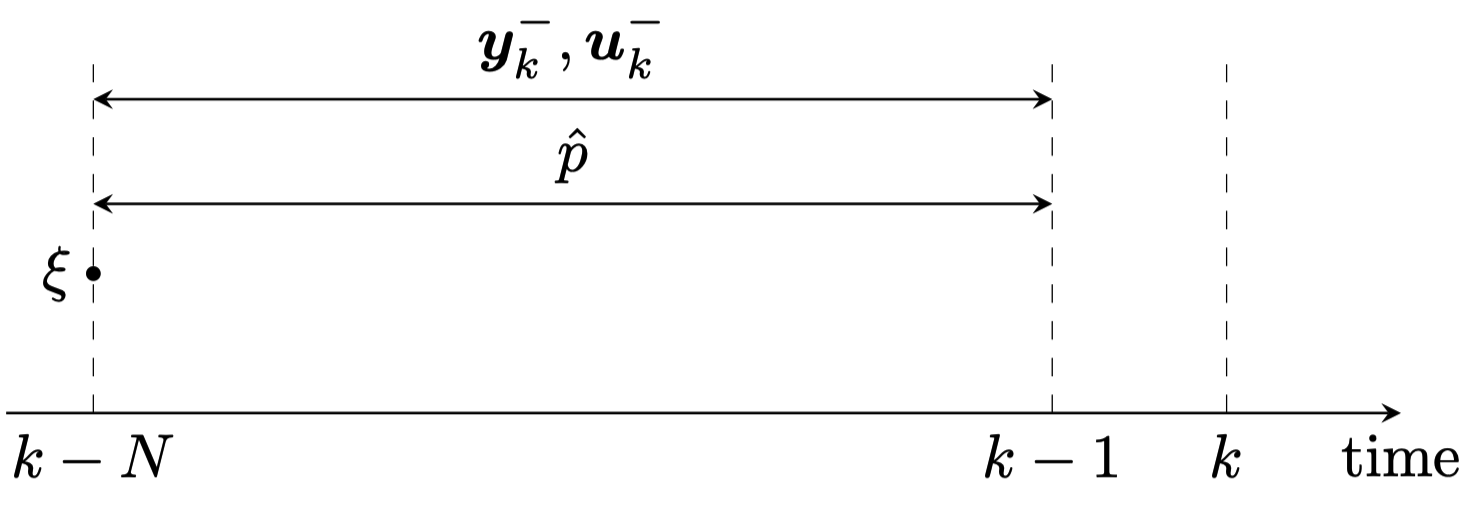}
\end{center}
\caption{Time positioning of some of the involved variables.} \label{positioningxi}
\end{figure}
\noindent and assuming that the input is measured so that $\bm u^-_k$ can be viewed as a part of $\bm y_k^-$, the last relation can be rewritten shortly as follows:
\begin{equation}
\bm e_k:=\mathcal E(\xi, \bm y_k^-, \hat p) \label{erreur}
\end{equation}
In order to  explicitly acknowledge the presence of  measurement errors and parametric uncertainties, we follow here the suggestion made in \cite{ALAMIR20091052} which amounts at introducing a dead-zone when evaluating the output prediction error, namely, given an error profile $\bm e^{[i]}$ on the output component $i$, we define the uncertainty-aware distance-to-zero $d$ as follows (in which $\lfloor r\rfloor_+=\max\{0, r\}$):
\begin{equation}
d(\bm e^{[i]},\zeta_i):=\sum_{k=1}^N\Big\lfloor\vert \dfrac{1}{k}\sum_{j=1}^{k}e_j^{[i]}\vert -[N/k]\zeta_i\Bigr\rfloor_+^r \label{defdeadzonei}
\end{equation} 
{\color{black} where $r$ is an integer that shape the penalty outside the dead-zone. $r=1$ is used in the numerical investigation section.}
Note that in the above expression, for each component $i$ of the output vector, $\zeta_{i}$ is the size of the dead zone which acknowledges that when the {\bf  average of the output prediction error} is below some threshold, this can be simply due to the measurement noise and parameter uncertainties and can therefore be discarded (clipped to $0$). Note that since each sensor might have different noise characteristics, it is necessary that different dead zone parameters $\zeta_i, i=1,\dots, n_y$ can be allowed, namely, one for each component of the measured output. \\
{\color{black} 
\begin{remark}[Rational behind dead-zone]\label{remzeta}\ \\
\textit{
In order to better understand the rational behind the use of the dead-zone size parameters $\zeta_i$, it is worth anticipating the forthcoming development to mention that the size $\zeta$ of this zone will be optimized so that the statistically guaranteed upper bound on the estimation error on the observation target is minimized. When this optimization step leads to a non vanishing size of the dead-zone, this simply means that  there is no smaller values that would be consistent, namely for which the correct value of the observation target can still be eligible regardless of the possible realizations of the measurement noise and the uncertain parameters. Consequently, the resulting non-zero radius of the deadzone would be an indicator of indistinguishability issue. In other words,  non vanishing resulting dead-zone sizes inform us that given the level of noise and uncertainties, the price to pay to keep eligible the true value of the targeted variable is to include other possible values that explains equally well the previously acquired measurements.}
\end{remark}}
\ \\
Based on this definition, the overall output prediction penalty is defined by:
\begin{equation}
d(\bm e, \zeta) := \sum_{i=1}^{n_y}d(\bm e^{[i]}, \zeta_i) \label{distance}
\end{equation}
Gathering together equations (\ref{distance}) and (\ref{erreur}), it is possible to define at each instant $k$ a cost function:
\begin{equation}
J(\xi, \hat p \ \vert\  \bm y^-_k, \zeta):=d\Bigl(\mathcal E(\xi, \bm y_k^-, \hat p), \zeta\Bigr) \label{khkjhkjhkjho}
\end{equation}
The relevance of this cost function can be stated as follows:
\\ \ \\ 
\begin{minipage}{0.02\textwidth}
\color{Brown} \rule{1mm}{30mm}
\end{minipage}
\begin{minipage}{0.45\textwidth}
\color{Brown} Given a vector of past measurements $\bm y^-_k$ (including the input profile) and a dead zone sizes vector $\zeta$, 
any pair $(\xi,\hat p)$ that induces  $J(\xi, \hat p \ \vert\  \bm y^-_k, \zeta)=0$ represents a possible explanation of the previous measurements $\bm y^-_k$  in which, $\xi$ is a possible value of the state at instant $k-N$ while $\hat p$ is a possible realization of the unknown parameter vector $p$.
\end{minipage}
\\ \ \\ \ \\
This cost function might be used in an extended MHE that first solves the optimization problem 
\begin{equation}
(\xi^\star, p^\star) = \mbox{arg}\min_{(\xi,p) \in \mathbb X\times \mathbb P} J(\xi,p\ \vert\  \bm y^-_k, \zeta)\label{MHOoptextended}
\end{equation}
and then uses the solution to reconstruct the observation target according to
\begin{equation}
\hat z_k = T\Bigl(X_N(\xi^\star,\bm u^-_k,p^\star), p^\star\Bigr)\label{defdexhatextended}
\end{equation}
The {\em raison d'\^{e}tre} of the dead zone is to make the cost function $J$ equal to $0$ when taken at the pair $(\xi,p)$ that issued the output measurment $\bm y=Y(\xi, \bm u, p)+\bm \nu$. For this to hold, the dead zone size $\zeta$ should be taken high enough. This leads to the following concept: \\ 
\hrule \vskip 2mm
\begin{definition}\label{defconsistency}[Dead-Zones Consistency] 
\color{Brown}
The vector of dead zones sizes $\zeta\in \mathbb{R}^{n_y}_+$ is said to be {\bf consistent} if for any admissible state $x\in \mathbb X$, any admissible control profile $\bm u\in \mathbb U^N$ and any possible realization of the vector of parameters $p\in \mathbb P$, the following equality is satisfied for any realization $\bm\nu\in \mathbb V^N$ of the noise profile:
\begin{equation}
J(x,p\ \vert \ \bm{Y}(x,\bm u, p)+\bm \nu, \zeta)=0 \label{J0}
\end{equation}
{\color{black} which simply means that the correct pair $(x,p)$ corresponds to a vanishing computed cost for any noise and any possible realization of the uncertain parameter vector $p$}.
\end{definition}
\vskip 1mm \hrule \vskip 2mm
\ \\
Let us recall that the output measurement vector contains the input information so that given $x$, $p$ and the output, one can simulate the system to produce a predicted trajectory of the system. Note also that the dead zones can always be made consistent by using sufficiently high values, but too high values would lead to  artificially big non distinguishability issue as it is discussed in the sequel. 
\section{The $\epsilon$-observability of an observation target} \label{problemstatement}
\noindent In the remainder of this contribution, we shall use the following notation:
\begin{equation}
q:= \begin{bmatrix}
 x\cr p\cr \bm u\cr \bm \nu
\end{bmatrix}=: \begin{bmatrix}
 q_x\cr q_p\cr q_{\bm u}\cr 
 q_{\bm \nu}
\end{bmatrix}  \in \mathbb Q\subset \mathbb X\times \mathbb P\times \mathbb U^N\times \mathbb V^N
\end{equation}
where the second equality is to be considered component-wise, namely, $x=q_x$, $p=q_p$ and so on. This enables to rewrite (\ref{J0}) in the following more condensed form:
\begin{equation}
(\forall q\in \mathbb Q)\quad J_1(q,\zeta):=J(q_x, q_p\ \vert \bm Y(q_x, q_{\bm u}, q_p)+q_{\bm \nu}, \zeta)=0 \label{defdeJw}
\end{equation}
Note that each element $q\in \mathbb Q$ completely defines a simulation scenario with an associated measurement noise profile. 

The following definition associates elements of $\mathbb Q$ that share the same exogenous information, namely the control input profile and the measurement noise realization: \\ 
\vskip 1mm \hrule \vskip 2mm
\begin{definition}[Comparable pairs]\label{defcontext} 
\color{Brown}
We shall say that two elements $q^{(1)}, q^{(2)} \in \mathbb Q$ are {\bf comparable} if and only if they share the components $\bm u$ and $\bm \nu$. This is denoted as follows:
\begin{equation}
q^{(1)} \bowtie q^{(2)}\quad \Leftrightarrow \quad (q_{\bm u}^{(1)}=q_{\bm u}^{(2)})\ \mbox{and}\ (q_{\bm \nu}^{(1)}=q_{\bm \nu}^{(2)})
\end{equation}
Such two elements obviously define two simulations that can differ only by the initial state and/or the vector of parameters while the input profile and the measurement noise are the same. 
\end{definition}
\vskip 1mm \hrule \vskip 2mm
\ \\ 
The second definition introduces the indistinguishability relationship on the set $\mathbb Q$ :\\ 
\vskip 1mm \hrule \vskip 2mm
\begin{definition}[indistinguishable pairs]\label{defequiv}
\color{Brown}
We shall say that an element $q^{(2)}$ is indistinguishable from $q^{(1)} \in \mathbb Q$ (with the notation $q^{(1)}\equiv q^{(2)} $)  if and only if it induces $0$ output prediction error cost when using as measurement the noisy output generated by $q^{(1)}$, namely
\begin{equation}
J(q_x^{(2)}, q_p^{(2)}\ \vert \bm Y(q_x^{(1)}, q_{\bm u}^{(1)}, w_p^{(1)})+q_{\bm \nu}^{(1)}, \zeta)=0 \label{eq26}
\end{equation}
meaning that $q_x^{(2)}$ and $q_p^{(2)}$ might as well explain the output induced by $q_x^{(1)}$ and $q_p^{(1)}$ for the same exogenous inputs $q_{\bm u}^{(1)}$ and $q_{\bm \nu}^{(1)}$.
\end{definition}
\vskip 1mm \hrule \vskip 2mm
\ \\
When a moving horizon estimator operates on some measurement data, it always works on comparable pairs of $q$ values {\color{black} (unless the MHE formulation involves the reconstruction of the measurement noise which is not the setting used here)}. Indeed, MHE tries to find a possibly optimal pair of initial states and parameter vector that can explain a given past measurements (including control input) that are obtained for a specific and unique realization of the measurement noise. The ideal situation is the one where no comparable but distinct values of $q$ can be indistinguishable. In this case, the only possible minimum of the above defined optimization problem is the one that involves the correct values of the initial state and the vector of parameters. However, all the previous discussion that justifies the current contribution tried to recall that in realistic situations, this is never the case. 
\\ \ \\ 
Note however that in the above ideal situation, one obtains perfect estimation of the state {\sc  and} the parameter vector which might be unnecessary in the case where only observation-target variable needs to be reconstructed. That is the reason why the following definition is introduced:\\ 
\vskip 1mm \hrule \vskip 2mm
\begin{definition}[$\epsilon$-observability]\label{defxdisting}
\color{Brown}
Given an observation-target $z=T(x,p)$, we shall say that $z$ is $\epsilon$-observable on $\mathbb Q$ if and only if {\color{black} 
\begin{enumerate}
\item There exists a dead-zone size vector $\zeta\in \mathbb{R}^{n_y}_+$ that is consistent over $\mathbb Q$ in the sense of Definition \ref{defconsistency},
\item The following implication holds for any pair $(q^{(1)}, q^{(2)})\in \mathbb Q^2$:
\end{enumerate}
\begin{equation}
\Bigl(q^{(1)}  \bowtie q^{(2)}\Bigr)\ \mbox{\sc and}\  \Bigl(q^{(1)}\equiv q^{(2)}\Bigr)\ \Rightarrow \ \|q_z^{(1)}-q_z^{(2)}\|\le  \epsilon \label{epsilonobs}
\end{equation}
where $q_z:=T(q_x, q_p)$ is the observation-target variable associated to $q$. In other words, only pairs with $\epsilon$-distant observable targets can be both comparable and indistinguishable.}
\end{definition}
\vskip 1mm \hrule \vskip 2mm
\ \\
The condition (\ref{epsilonobs}) can be written equivalently as follows thanks to (\ref{eq26}):
\begin{equation}
\begin{array}{c}
\Bigl(q^{(1)}  \bowtie q^{(2)}\Bigr)\quad \mbox{\sc and}\quad \Bigl(\|q_z^{(1)}-q_z^{(2)}\|> \epsilon\Bigr) \\
	\Downarrow \\
	J(q_x^{(2)}, q_p^{(2)}\ \vert \bm Y(q_x^{(1)}, q_{\bm u}^{(1)}, q_p^{(1)})+q_{\bm \nu}^{(1)}, \zeta)\neq 0
\end{array} \label{eq28}
\end{equation}
We shall rewrite this last implication in a more practical form using the following steps:
\begin{itemize}
\item Let us pick some $q^{(1)}=q$ that is an arbitrarily chosen element inside $\mathbb Q$
\item Any comparable $q^{(2)}$ (an element such that $q^{(1)} \bowtie q^{(2)}$) takes necessarily the form:
\begin{equation}
q^{(2)} := \begin{bmatrix}
\xi\cr p\cr q_{\bm u}\cr q_{\bm\nu}
\end{bmatrix} 
\end{equation}
for some $(\xi,p)\in \mathbb Z:=\mathbb X\times \mathbb P$.
\item Now in order for the second condition in the left hand side of (\ref{eq28}) to hold, (namely $\vert q_z^{(1)}-q_z^{(2)}\vert \ge \epsilon$), we shall restrict $(\xi,p)$ to the set $\bar{\mathbb Z}_\epsilon(q)$ which is the complement of $\mathbb Z_\epsilon(q)$ defined by:
\begin{equation*}
\mathbb Z_\epsilon(q) := \Bigl\{(\xi,p)\in \mathbb X\times \mathbb P \ \mbox{\rm s.t.}\ \|T(\xi,p)-T(q_x,q_p)\|\le \epsilon\Bigr\}
\end{equation*}
\end{itemize}
Using the above notation, the implication (\ref{eq28}) can be written in the following more compact form: 
\begin{align}
&\forall q\in \mathbb Q \quad \forall (\xi, p)\in\bar{\mathbb Z}_\epsilon(q) \nonumber  \\ 
&J(\xi, p\ \vert\ \bm Y(q_x,q_{\bm u}, q_p)+q_{\bm\nu}, \zeta)\neq 0 \label{eq29}
\end{align}
and introducing the notation:
\begin{align}
w&:= \begin{bmatrix}
 q\cr \xi\cr p
\end{bmatrix} \in \mathbb W:=\mathbb Q\times \mathbb X\times \mathbb P \label{defdewwww} \\
\bar{\mathbb W}(\epsilon)&:=\Bigl\{w=\begin{bmatrix}
 q\cr \xi\cr p
\end{bmatrix}\ \vert \ (q,\xi,p)\in \mathbb Q\times\bar{\mathbb Z}_\epsilon(q)\Bigr\} \label{defdeWeps}\\
J_2(w,\zeta)&:=J(\xi, p\ \vert\ \bm Y(q_x,q_{\bm u}, q_p)+q_{\bm\nu}, \zeta) \label{defQepsJ}
\end{align}
one gets the following result: \\ 
\vskip 1mm \hrule \vskip 2mm
\begin{proposition} [First formulation]\label{prop1}
An observation-target variable $z$ is $\epsilon$-observable on $\mathbb Q$ using the deadzone vector $\zeta$ if and only if the following conditions hold true:
\begin{enumerate}
\item The dead-zone sizes vector is consistent in the sense of Definition \ref{defconsistency}, namely:
\begin{equation}
(\forall q\in \mathbb Q)\qquad J_1(q,\zeta)=0 \label{C1}
\end{equation}
\item The $\epsilon$-distinguishability property holds true, namely
\begin{equation}
(\forall w\in \bar{\mathbb W}(\epsilon)) \quad J_2(w,\zeta)\neq 0 \label{C2}
\end{equation}	
\end{enumerate}
where $\bar{\mathbb W}(\epsilon)$ is defined by (\ref{defdeWeps}) while  $J_1$ and $J_2$ are defined by (\ref{defdeJw}) and (\ref{defQepsJ}) respectively. 
\end{proposition}
\vskip 1mm \hrule \vskip 2mm
\ \\
Recall that the first condition of Proposition \ref{prop1} ensures that the correct pair leads to an output prediction cost that is equal to $0$ and is therefore always an admissible solution to the optimization problem  while the second condition guarantees that there are no indistinguishable pairs whose observation-target variables are distant by more that $\epsilon$. 
\\ 
Note that the condition (\ref{C1}) can also be written using the notation $w_q$ that extracts the first vector $q$ in the vector $w$ [see (\ref{defdewwww})] so that one can write the condition (\ref{C1}) as follows:
\begin{equation}
(\forall w\in \bar{\mathbb W}(\epsilon))\qquad J_1(w_q,\zeta)=0 \label{C1bis}
\end{equation}
This enables to regroup the two conditions (\ref{C1}) and (\ref{C2}) in a single condition that involves the parameter $w$, namely:
\begin{align}
&(\forall w\in \bar{\mathbb W}(\epsilon))\nonumber \\ 
&C(w,\zeta):= \left\{ 
\begin{array}{ll}
  0& \mbox{\rm if $J_1(w_q,\zeta)=0\quad \mbox{\rm and}\quad J_2(w,\zeta)\neq 0$}\\
 1 & \mbox{otherwise}
\end{array}
\right.
\end{align}
This enables Proposition \ref{prop1} to be reformulated in a more compact form that will be more convenient for the formulation of the probabilistic certification step: 
\vskip 1mm \hrule \vskip 2mm
\begin{proposition}[Second formulation]\label{prop2}
An observation target variable $z$ is $\epsilon$-observable on $\mathbb W$ with the dead-zone vector $\zeta$ if and only if the following condition holds true:
\begin{equation}
(\forall w\in \bar{\mathbb W}(\epsilon))\qquad C(w,\zeta)=0\label{C1bbis}
\end{equation}
\end{proposition}
\vskip 1mm \hrule \vskip 2mm
\ \\
For technical reasons, we need to perform a last transformation by observing that the condition (\ref{C2}) that concerns only those values of $w$ that belong to $\bar{\mathbb W}(\epsilon)$ can be transformed into a condition on all possible values of $w\in \mathbb W:=\mathbb Q\times \mathbb X\times \mathbb P$ by writing
\begin{equation*}
(\forall w\in \mathbb W)\qquad g(w,\epsilon,\zeta)=0 
\end{equation*}
where 
\begin{align}
&g(w,\epsilon,\zeta):= \nonumber \\ 
&\left\{ 
\begin{array}{ll}
 0& \mbox{\rm if $J_1(w_q,\zeta)=0$ and $w\notin \bar{\mathbb W}(\epsilon)$}\\
 0& \mbox{\rm if $J_1(w_q,\zeta)=0$ and $\Bigl(w\in \bar{\mathbb W}(\epsilon)\ \mbox{and }\ J_2(w,\zeta)\neq 0\Bigr)$} \\
 1 & \mbox{otherwise}
\end{array}
\right. \label{defdeg}
\end{align}
Note that in above formulation the pair defined by 
\begin{equation}
\theta := \begin{bmatrix}
 \epsilon\cr \zeta 
\end{bmatrix}  \label{defdetheta}
\end{equation}
is viewed as a design parameter vector for the probabilistic certification setting. The same notation $\theta_\epsilon=\epsilon$ and $\theta_\zeta=\zeta$ will also be used to invoke the individual components of this design vector. The above notation enables to formulate the final form of the $\epsilon$-observability formulation: \\ 
\vskip 1mm \hrule \vskip 2mm
\begin{proposition}[Final Formulation]\label{prop3}
The observation target is $\epsilon$-observable on $\mathbb W$ with the dead-zone vector $\zeta$ if and only if the pair $\theta=(\epsilon,\zeta)$ satisfies the following condition:
\begin{equation}
(\forall w\in \mathbb W) \qquad g(w,\theta)=0 \label{CFT}
\end{equation}
\end{proposition}
\vskip 1mm \hrule \vskip 2mm
\ \\
The condition (\ref{CFT}) is called a {\bf robust constraints satisfaction condition} as the satisfaction of the constraint $g(w,\theta)=0$ is required for {\bf all} possible realizations of the argument $w$ and this, regardless of its probability of occurrence. \\ \ \\ Such a condition shows two major drawbacks:
\begin{enumerate}
\item The first is that such formulation leads to very pessimistic results since the impossibility to meet the robust constraint satisfaction condition might be due to one single very unlikely value (or a set of values) of the parameter $w$.\\
\item Even if required, checking the robust satisfaction constraint property is extremely complex (impossible actually) since one needs to check the satisfaction of the condition for all possible values of $w$. Now one can argue that there is no need to check all the values since instead one can {\em simply} check the following condition:
\begin{equation}
\max_{w\in \mathbb W} \vert g(w,\theta)\vert =0 \label{optalternative}
\end{equation}
which is mathematically equivalent to the previous condition while it does not necessarily mean that all the values should be individually checked. \\ \ \\ The problem with this alternative is that while it might be possible for convex problems, it remains a hard-to-check assertion in the general case where one can scarcely be sure that the optimizer does identify the global maximum. The problem is therefore replaced by the one consisting in asserting that the numerically obtained maximum in (\ref{optalternative}) is really a {\sc  global} maximum which is obviously as hard to decide as the original problem for general non convex settings. \\
\end{enumerate}
This discussion leads to the concept introduced in the following section.
\section{The $\eta$-almost $\epsilon$-observability} \label{secproba}
\noindent In order to avoid these difficulties and to come with a realistic assessment of the observability, the following less stringent concept of observability is introduced: \\ 
\vskip 1mm \hrule \vskip 2mm
\begin{definition}\label{deffrt}[Almost $\epsilon$-observability] 
\color{Brown}
Given a predefined observation-target variable $z$, given a small $\eta\in (0,1)$ and assuming some probability distribution $\mathcal W$ that governs the dispersion of the context parameter $w$ we say that the observation target $z$ is $\eta$-almost $\epsilon$-observable if and only if there is a design parameter $\theta=(\epsilon,\zeta)$ such that the following inequality holds true:
\begin{equation}
\mbox{\rm Pr}_{\mathcal W}\Bigl[g(\cdot,\theta)\neq 0\Bigr]\le \eta \label{prob1}
\end{equation} 
meaning that the $\epsilon$-observability condition (\ref{CFT}) is satisfied with a high probability (1-$\eta$).
\end{definition}
\vskip 1mm \hrule \vskip 2mm
\ \\
There is a variety of probabilisitic certification settings that have been developed along the past recent years (see \cite{Calafiore2006,alamo2009,Calafiore2010,VIDYASAGAR20011515} and the references therein for an overview). For the sake of conciseness and in order to focus on the main ideas regarding the certification of partial observability, only one version among various possible settings is pursued. Alternative developments can be undertaken later if appropriate without modification in the main message and philosophy. \\ \ \\ 
To do so, the specific version of the probabilistic certification settings used in the sequel is first recalled in the next section. 
\subsection{Probabilistic certification framework}
\noindent Although the formulation of (\ref{prob1}) is less stringent that (\ref{CFT}), it is still difficult to manipulate as the probability of a nonlinear map depending on a vector of stochastic variables remains impossible to compute. That is why probabilistic certification approach introduces a {\sc  second step} consisting in approximating (\ref{prob1}) via numerical averaging over a high number $N_s(\eta,\delta,m)$ of sampled realizations of the stochastic variable $w$. This number $N_s(\eta,\delta,m)$ depends on: \\
\begin{itemize}
\item The certification precision parameter $\eta\in (0,1)$ introduced above [see (\ref{prob1})].\\
\item The certification confidence parameter $\delta\in (0,1)$ which defines the degree of confidence with which the certification statement can be delivered, namely, the probability that the statement holds is greater than $1-\delta$.\\
\item $m\in \mathbb N_*$ is the maximum number of admissible violations of the constraints among the $N_s(\eta,\delta,m)$ drawn samples. This simply means that when testing the condition $g(w^{(i)},\theta)=0$ on a high number $N_s(\delta,\eta,m)$ of samples $\{w^{(i)}\}_{i=1}^{N_s(\eta,\delta,m)}$ that are drawn (using $\mathcal W$) no more than $m$ samples lead to constraint violation. This is equivalent to write:
\begin{equation}
\sum_{i=1}^{N_s(\delta,\eta,m)} g(w^{(i)},\theta)\le m \label{enfinprob}
\end{equation}
as $g=0$ when the constraint are satisfied while $g=1$ when the constraint is violated.
\end{itemize}
Based on the above notation, the probabilistic certification of the $\epsilon$-observability concept is given by the following proposition \cite{ALAMO20051035}:\\ 
\vskip 1mm \hrule \vskip 2mm
\begin{proposition}[Certification of $\epsilon$-observability] \label{propprinc}
Consider the following setting's components:
\begin{itemize}
\item A given discrete set $\Theta\subset \mathbb{R}^{2}_+$ containing $n_\Theta$ values of the design parameter $\theta:=(\epsilon,\zeta)$.
\item A certification confidence parameter $\delta\in (0,1)$
\item A certification precision parameter $\eta\in (0,1)$
\item  A maximum number of failures $m\in \mathbb N_*$
\item An integer $N_s$ satisfying:
\begin{equation}
N_s\ge \dfrac{1}{\eta}\Bigl[m+\ln(\dfrac{n_\Theta}{\delta})+(2m\ln(\dfrac{n_\Theta}{\delta}))^{\frac{1}{2}}\Bigr] \label{ineqN}
\end{equation}
\item $N_s$ realizations $\{w^{(i)}\}_{i=1}^{N_s}$ of the vector $w$ drawn randomly according to $\mathcal W$ inside $\mathbb W$.
\end{itemize}
If there is an element $\theta^\star=(\epsilon^\star,\zeta^\star)\in \Theta$ such that the following inequality holds true:
\begin{equation}
\sum_{i=1}^{N_s} g(w^{(i)},\theta^\star)\le m \label{enfinprobprop}
\end{equation}
Then the condition 
\begin{equation}
\mbox{\rm Pr}_{\mathcal W}\Bigl(g(\cdot,\theta^\star)\neq 0\Bigr)\le \eta \label{prob1star}
\end{equation}
is satisfied with a probability greater than $1-\delta$. 
\\ \ \\ 
In other words, if the condition (\ref{enfinprobprop}) holds for $N_s$ satisfying (\ref{ineqN}) then the observation target is $\eta$-almost $\epsilon^\star$-observable in the sense of Definition \ref{deffrt}.
\end{proposition}
\vskip 1mm \hrule \vskip 2mm
Based on the above discussion, the randomized optimization framework amounts at solving an optimization problem of the form:
\begin{equation}
\min_{\theta\in \Theta} \Bigl[\mbox{\rm cost}(\theta)\Bigr] \qquad \mbox{\rm under}\quad \sum_{i=1}^{N_s} g(w^{(i)},\theta)\le m\label{enfinprobpropjh}
\end{equation}
where $N_s$ is defined by (\ref{ineqN}) in which $m$ is the number of admissible constraint-violating scenarios. \ \\ 
In our case, the cost function in (\ref{enfinprobpropjh}) that is defined in terms of the decision variable $\theta:=(\epsilon,\zeta)$ is obviously given by:
 $$\mbox{\rm cost}(\theta):=\epsilon$$
 since the objective is to get a certification results with the lowest state estimation error on the observation target variable while $\zeta$ is simply a design parameter of the MHE. Before we examine a specific illustrative example, some general comments and discussion is proposed relative to different aspects of the implementation.
\subsection{Discussion regarding the choice of the statistics of the random sampling}
\noindent When applying the above framework to a specific problem, the choice of the probability distribution $\mathcal W$ that is to be used in drawing the $N_s$ realization is a quite difficult one. Recall that $\mathcal W$ is the probability distribution that defines the statistics of occurrence of elements inside the set
$$\mathbb W= \underbrace{\Bigl(\mathbb X\times \mathbb P\times \mathbb U^N\times \mathbb V^N\Bigr)}_{\mathbb Q}\times \mathbb X\times \mathbb P$$
As far as state samples ($\in \mathbb X$) are concerned, this distribution should reflect the effectively encountered one when the state observation framework is applied to the system. For instance, when the system is controlled in order to follow some collection of desired set-points, it is obvious that the state will be more frequently present in the neighborhood of this specific set of values so that these regions should be more heavily present in a relevant sampling of the component of $q_x$ of $q=w_q$. The same can be said about the distribution of the components of the control profile $q_{\bm u}$ of $q=w_q$. \\ \ \\ 
Nevertheless, the system needs to make transitions between these steady states and observability is needed during these transients in order to certify their success in a sufficiently high confidence rate. This means that a certification of the observability that would be based on a probability distribution of states that considers the transient states as instances with very low probability would be problematic (since too optimistic with regards to the very reason for which observability needs to be assessed with a high confidence rate). \\ \ \\ 
On the other hand, using uniform distributions might include regions of the state space that are never visited by the system for many reasons. This means that including them with non vanishing probability of occurrence might lead to pessimistic results regarding the observability if observability does not hold at these regions, or to an optimistic results if these regions correspond to high observability patterns. \\ \ \\ 
As for the statistics of the parameter vector $p$, this is obviously a problem-dependent choice that relies on a deeper understanding of the reason for parameters dispersion. For instance, if it is about production-related dispersion, it might be reasonable to use Gaussian around the nominal manufacturing values. If it is about biological model's parameters that depend on some characteristics (age, gender, weight, etc), then certification can be done for each category apart or mixed weighted multi-gaussian distribution can be used depending on the statistics of these characteristics among the population under interest. \\ \ \\ 
It comes out from the previous discussion that there is probably not a single perfect choice of $\mathcal W$ and that this question probably deserves a dedicated study on its own in every specific case. As far as this paper is concerned, several choices are tested and compared in order to show the sensitivity of the result to these choices. 
It is worth underlying however that despite of the difficulty in choosing the sampling set, any setting that leads to certification-related difficulty does reveal that observability might be an issue. 
\subsection{The investigated statistics}
\noindent In order to completely define the statistics $\mathcal W$, we need to define the sampling rules of all the sets involved in the definition of $\mathbb W$, namely, $\mathbb X$, $\mathbb P$, $\mathbb U$ and $\mathbb V$. The choices used in the numerical investigation of this paper are defined as follows. 
\begin{description}
\item[$\mathbb X\qquad$] A uniform distribution over a hyperbox $\mathbb X$\\
\item[$\mathbb P\qquad$] Two possibilities are investigated, namely:\\
\begin{enumerate}
\item  A uniform distribution over $\mathbb P$
\item A Gaussian distribution around a nominal value.\\
\end{enumerate}
\item[$\mathbb U^N\quad$] Two possibilities are investigated, namely:
\begin{enumerate}
\item A uniform distribution over $\mathbb U^N$
\item A random choice consisting in sequences (elements of $\mathbb U$) of the form ($i\in \{1,\dots, N\})$:  
\begin{equation}
u_i=\mbox{\rm Sat}_\mathbb{U}\Biggl[\sum_{j=1}^{n_f}\beta_j\sin(\dfrac{2j\pi(i\tau)}{N\tau}+\varphi_j)\Biggr]\label{defdesinu}
\end{equation}
where $\mbox{\rm Sat}_\mathbb{U}$ is the projection map on $\mathbb U$ while the coefficients $\beta_j\in \mathbb{R}^{n_u}$ and the phases $\varphi_j$ for $j\in \{1,\dots,n_f\}$ are uniformly randomly selected in $[0,\bar\beta]$ and $[0,2\pi]$ respectively. This sequence can be as rich as desired (by taking a high value of $n_f$) to represent any possible behavior of the feedback law. 
\end{enumerate}
\item[$\mathbb V^N\quad$] Gaussian white noise (with different variances) are used to represent measurement noise. 
\end{description}
\subsection{The design set}
\noindent For each considered configurations of sampling statistics, $N_s$ samples are drawn with $N_s$ satisfying the inequality (\ref{ineqN}). The probabilistic certification requires finding $\theta^\star=(\epsilon^*,\zeta^*)$ such that the inequality (\ref{enfinprobprop}) holds over the set of sampled $w^{(i)}$, except at most $m$ instances, where $\theta^\star$ belongs to a before hand defined discrete set $\Theta$ of cardinality $n_\Theta$. \\ \ \\ 
In what follows the structure of the set of design parameter $\Theta$ is taken of the form:
\begin{equation}
\Theta := \mathbb L(\underline\sigma_\epsilon, \bar\sigma_\epsilon, n_\epsilon)\times \mathbb L(\underline\sigma_\zeta, \bar\sigma_\zeta, n_\zeta) \label{dessetparam}
\end{equation}
where $\mathbb L(\underline\sigma,\bar\sigma,n)$ is the set of $n$ logarithmically uniformly spaced numbers, namely\footnote{As an example, $\mathbb L(-2,0,5)\approx \{0.01, 0.032, 0.1, 0.32, 1.        ]\}$}:
\begin{align*}
&\mathbb L(\underline\sigma,\bar\sigma,n)= \mbox{\rm logspace}(\underline\sigma,\bar\sigma,n)\nonumber \\ 
&:=\Bigl\{10^{r_i}\quad\vert\quad  r_i=\underline \sigma+\dfrac{(\bar\sigma-\underline \sigma )i}{n-1}\quad i\in \{0,\dots,n-1\}\Bigr\}
\end{align*}
This obviously leads to a cardinality $n_\Theta=n_\epsilon n_\zeta$
\subsection{Implementation and complexity analysis} \label{seccomplexity}
It is shown shortly that the specific observability problem leads to a specific complexity analysis but in a more general settings of (\ref{enfinprobpropjh}), since we are using a discrete set $\Theta$ of admissible values of  $\theta$, the worst case analysis involves a number of simulations that is equal to $N_sn_\Theta=N_sn_\epsilon n_\zeta$ simulations which corresponds to an exhaustive search. This can be quite heavy if the simulation of the system is cumbersome. But this is quite rarely the case for systems that are analytically given and which involve a reasonable state space dimension. Examples of such realistic exhaustive search settings can be found in \cite{ALAMIR201559,PFLAUM2017596}. Such a brute force approach can be avoided by using standard iteration algorithms on discrete-sets although the worst case analysis remains dependent on the product $N_sn_\epsilon n_\zeta$. \\ \ \\ 
For the specific certification problem of almost $\epsilon$-observability, it turns out that the structure of the constraint function $g(w,\theta)$ takes the following form:
\begin{equation}
g(w,\theta)=G(\bm e(w),\theta)
\end{equation}
where $\bm e$ is the output prediction error before dead-zone clipping while $G$ is a very cheap map that mainly consists in clipping followed by conditional summation over the resulting profiles. Consequently, an exhaustive search is more affordable than in the general case since the  computation consists in three successive steps which are:
\begin{enumerate}
\item  Simulating the $N_s$ scenarios using the set $\{w^{(i)}\}_{i=1}^{N_s}$ sampled instances to get the set $$\{\bm e^{(i)}:=\bm e(w^{(i)})\}_{i=1}^{N_s}$$
\item Computing for each $\theta\in \Theta$ the resulting cost and constraints for each $\bm e^{(i)}$ using $G(\bm e^{(i)},\theta)$
\item Select among all values of $\theta$ satisfying the constraints the one that corresponds to the lowest $\epsilon=\theta_1$.
\end{enumerate}
\ \\ This leads to a worst case complexity of ($N_s$ simulations + $n_\Theta$ evaluations of $G$) which corresponds to a much less complexity than in the general case. Moreover, by ordering the elements of $\Theta$ is ascending-in-$\theta_1$ order, one can stop as soon as the constraints is satisfied leading to generally less than total exhaustive search.  Despite of the above guidelines, more involved optimization schemes can be investigated but which are out of the scope of this contribution. \\ \ \\ 
The whole certification framework is summarized in Algorithm \ref{algo1}. 

\begin{algorithm}
\caption{The certification algorithm}\label{algo1}
\color{black}
\begin{algorithmic}[1]
\State {\bf Given:}
\begin{itemize}
\item The maps  $f$ and $h$ \Comment{[see (\ref{model}) and (\ref{defdeh}))}
\item $\eta, \delta, m$ \Comment{Certification parameters [see (\ref{ineqN})]}
\item $\mathbb X, \mathbb P, \mathbb U, \mathbb V $ \Comment{Working sets}
\item $\underline\sigma_\epsilon, \bar\sigma_\epsilon, \underline\sigma_\zeta, \bar\sigma_\zeta, n_\epsilon, n_\zeta$ \Comment{Design Set parameters (\ref{dessetparam})}
\item Sampling rule $\mathcal W$ on $\mathbb W=\mathbb Q\times \mathbb X\times \mathbb P$
\end{itemize}
\State $n_\Theta\leftarrow n_\epsilon\cdot n_\zeta$ \Comment{Cardinality of the design set $\Theta$}
\State Compute $\Theta:=\{\theta^{(j)}\}_{j=1}^{n_\Theta}$ by (\ref{dessetparam}) in alphabetic order
\State $N_s\leftarrow N_s(\delta,\eta,m,n_\Theta)$ by (\ref{ineqN}) \Comment{Number of scenarios}
\State Generate $N_s$ scenarios inputs $\{w^{(i)}\}_{i=1}^{N_s}$ using $\mathcal W$
\For {$i\in \{1,\dots, N_s\}$}
\State Compute $\bm e^{(i)}$ \Comment{non clipped error for $w^{(i)}$}
\EndFor
\State \texttt{success} $\leftarrow$ False \Comment{No solution $\theta^\star$ found yet}
\For{$j\in \{1,\dots, n_\Theta\}$}
\State \texttt{nb\_failures} $\leftarrow \sum_{i=1}^{N_s}g(w^{(i)}, \theta^{(j)})$
\If{\texttt{nb\_failures}$\le m$}
\State \texttt{success} $\leftarrow$ True \Comment{Optimal solution found}
\State $\theta^\star\leftarrow \theta^{(j)}$. {\bf Break}. \Comment{Stop the loop}
\EndIf
\EndFor
\State {\bf Output:} {\bf If} {\texttt{success}} {\bf return} $\theta^\star$ 
\end{algorithmic}
\end{algorithm}

\section{Illustrative example} \label{secNI}
\subsection{The dynamic system}
Consider the example of the nonlinear continuous stirred-tank reactor with parallel reaction \cite{Bailey1971}:
\begin{align*}
R&\rightarrow P_1\\
R&\rightarrow P_2\\
\end{align*}
that can be described by the following set of dimensionless energy and material balances:
\begin{align}
\dot x_1&=1-p_1x_1^2e^{-1/x_3}-p_2e^{-p_3/x_3}-x_1\\
\dot x_2&=p_1x-1^2e^{-1/x_3}-x_2\\
\dot x_3&=u-x_3
\end{align}
where $x_1$ and $x_2$ represent the concentrations of $R$ and $P_1$ while $x_3$ stands for the temperature of the mixture in the reactor. $P_2$ represents the waste product. This reactor is controlled by the manipulated variable $u\in \mathbb U:= [0.049, 0.449]$ (which represents the reaction temperature) in order to maximize the production of $x_2=P_1$. Note that the above dynamics involves $n_p=3$ parameters  $p_1$, $p_2$ and $p_3$ whose nominal values are commonly considered to be $p^{nom}=(10^4, 4\times 10^2, 0.55)$. It is assumed that $x_2$ is measured together with $u$ while $x_1$ and $x_3$ has to be estimated by the observer.  
\subsection{The framework setting}
\noindent The above settings is applied with the following definition of the subset $\mathbb X\subset \mathbb{R}^{3}$:
$$\mathbb X := [0,0.6]\times [0, 0.3] \times [0.05, 0.2]$$
which contains realistic evolutions of the state during realistic operational context. 
\begin{figure}[H]
\begin{center}
 \includegraphics[width=0.4\textwidth]{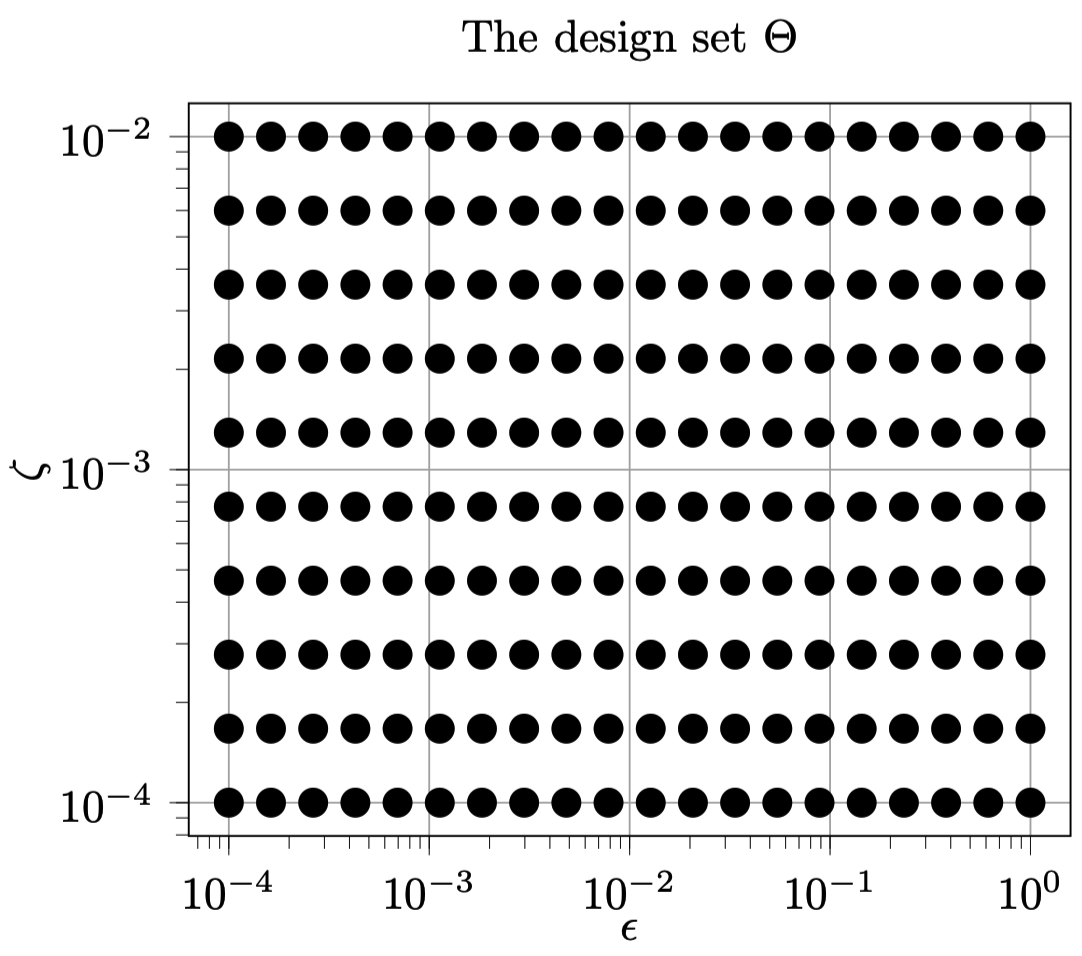}
\end{center}
\caption{The discrete design set $\Theta$ used in the probabilistic certification framework of proposition \ref{propprinc}.} \label{figsetTheta}	
\end{figure}
\ \\ 
The discrete design  set $\Theta$ containing the candidate values of the pair $(\epsilon,\zeta)$ is illustrated in Figure \ref{figsetTheta}. This set is of cardinality $n_\Theta=200$. \\ \ \\ 
Regarding the set of parameters that is sampled at each certification experiment, two alternatives are tested as mentioned above:
\begin{enumerate}
\item In the first, a uniform distribution is considered on the hyper-box:
\begin{equation}
\mbox{\rm (Uniform)}\quad \mathbb P := \Pi_{i=1}^{n_p} [(1-\rho)p^{nom}_i, (1+\rho)p_i^{nom}] \label{puniform}
\end{equation}
\item A random distribution centered at $p^{nom}$ such that each sample is given by
\begin{equation}
\mbox{\rm (Gaussian)}\quad p=\Pi_{i=1}^{n_p}(1+s_{td}\nu)p^{nom}_i \label{pgaussian}
\end{equation}
\end{enumerate}
where $\nu$ is a normal distribution centered at $0$. \\ \ \\ 
Regarding the control profiles that are sampled at each certification experiment, two alternatives are tested:\\
\begin{enumerate}
\item In the first, a uniform distribution is considered on the hyper-box $\mathbb U := [0.049, 0.449]^N$.\\
\item A truncated Fourier series with random coefficients [see (\ref{defdesinu})] with $n_f=10$. 
\end{enumerate}
The noise profile realizations have been generated using the uniform distribution over $[-\bar \nu, +\bar \nu]$. \\ \ \\ 
Finally, three observation-targets are investigated which are:
\begin{equation}
z_1=x\quad ;\quad z_2=x_1\quad;\quad z_3=x_3
\end{equation}
More precisely, applying the certification framework with the observation-target $z_1$ corresponds to standard  whole state observability while the use of $z_2$ [resp. $z_3$] corresponds to the cases where only the quality of the reconstruction  of $x_1$ [resp. $x_3$] matters. 
\subsection{Results}
\noindent Different aspects are successively examined in terms of their impact on the certification results. The results are shown through data frames in which the signification of the columns are as follows:
\begin{itemize}
\item  {\bf  eps1, eps2, eps3}. The certified reconstruction precision $\epsilon$ on $z_1$, $z_2$ and $z_3$ respectively. 
\item {\bf  zeta1, zeta2, zeta3}. The optimal computed  dead-zone sizes. \\ More precisely {\bf  (eps1, zeta1)}, {\bf (eps2, zeta2)} and {\bf (eps3, zeta3)} are the solutions of (\ref{enfinprobpropjh}) when the observation-target variable is $z_1$, $z_2$ or $z_3$ respectively. 
\item $\mathbf N$. The observation horizon.
\item {\bf  noise}. The noise level $\bar \nu$ mentioned above.
\item {\bf rho}. The size $\rho$ of the hyper-box as invoked in (\ref{puniform}) when the uniform distribution of the parameters is used (This corresponds to the value {\em uniform} in the column entitled {\bf p\_mode}).
\item {\bf  std\_p}. The parameter $s_{td}$ used in (\ref{pgaussian}) to define the gaussian distribution of the parameters around the nominal values. (This corresponds to the value {\em gaussian} in the column entitled {\bf p\_mode}).
\item {\bf  u\_mode}. The type of input used in the certification (can be {\em Fourier} or {\em rand}).  
\item {\bf  eta, delta, m}. The certification parameters, namely the certification precision $\eta$, confidence $\delta$ and the number $m$ of failures used in the probabilistic certification framework.
\end{itemize}
Before we examine the numerical results, it is important to clearly understand the meaning of the certifiable reconstruction precision $\epsilon$. This has to be understood as a probabilistically certified upper bound on the instantaneous estimation error. \\
\subsubsection{Impact of the observation horizon}
\begin{figure*}
\begin{center}
 \includegraphics[width=\textwidth]{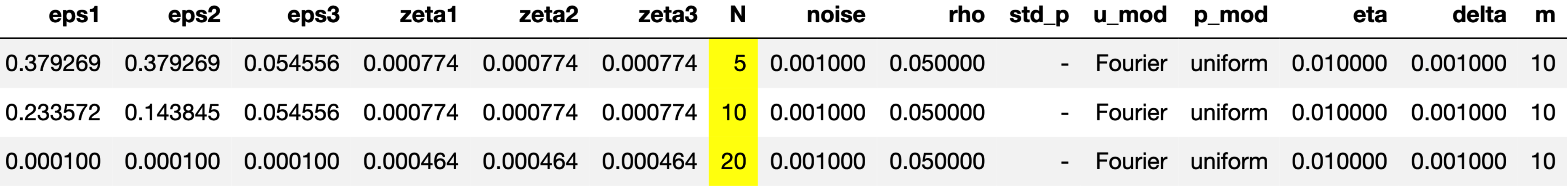}
\end{center}	
\caption{Impact of the observation horizon on observability.} \label{figdfN}
\end{figure*}

\ \\
\noindent Figure \ref{figdfN} shows the impact of the length of the observation horizon that is used to define the output prediction error cost. It is in particular shown that for the considered uncertainty setting, one needs to use $N=20$ in order to achieve the certification with the lowest values $\epsilon=10^{-4}$ considered in the design set $\Theta$ over the three observation-target variables $z_i$, $i=1,2,3$. Otherwise, indistinguishability might occur with quite high error values on the targeted indicators. Note that only three values of $N$ are studied here, lower values of $N\in [10,20]$ would have probably be sufficient to achieve the high precision certification results. 
\\ 
\subsubsection{Impact of the measurement noise}
\begin{figure*}
\begin{center}
 \includegraphics[width=\textwidth]{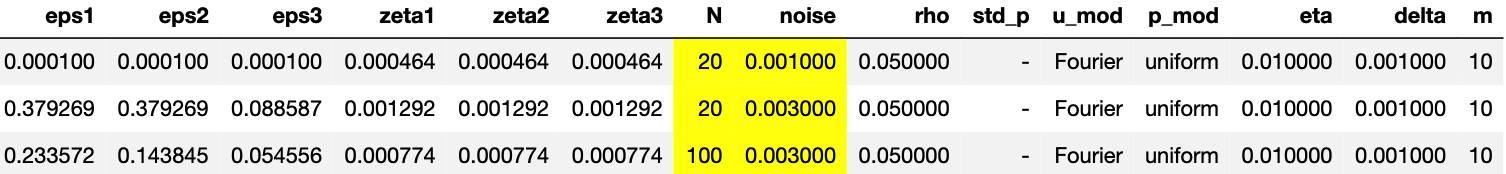}
\end{center}	
\caption{Impact of the measurement noise level on observability.} \label{figdfnoise}
\end{figure*}

\ \\ Starting from the last setting of Figure \ref{figdfN}, the noise level is increased from 0.001 to 0.003. This leads to a sensitive degradation in the certifiable reconstruction precision. Figure \ref{figdfnoise} shows that by increasing the observation horizon up to $N=100$ it is possible to recover the  levels of precision of the second setting of Figure \ref{figdfN} which was achievable with $N=10$ and the previous level (0.001) of the noise. This clearly shows that the higher the noise is the longer the observation horizon should be to achieve the same level of certifiable reconstruction precision. 
\\ 
\subsubsection{Impact of the parametric uncertainty level and statistics}
\begin{figure*}
\begin{center}
 \includegraphics[width=\textwidth]{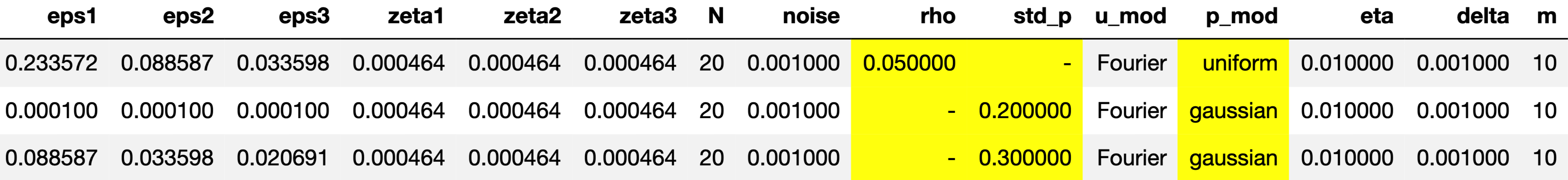}
\end{center}	
\caption{Impact of the parameter uncertainties level and distribution on observability.} \label{figdfp}
\end{figure*}

\ \\ In this investigation, the level and the type of parametric uncertainties are changed in order to evaluate their effect on the certifiable reconstruction precision. In particular, the  comparison between the first two lines of Figure \ref{figdfp} shows that the configuration with uniform distribution of the parameter vector given by (\ref{puniform}) with $\rho=0.05$ is more inconvenient to certifiable reconstruction precision than the gaussian distribution  given by (\ref{pgaussian}) and $s_{td}=0.2$. The third line shows what would be the certifiable reconstruction precisions on the three observation-targets when the last gaussian distribution parameter is increased from $0.2$ to $0.3$.
\ \\
\subsubsection{Impact of the input}
\begin{figure*}
\begin{center}
 \includegraphics[width=\textwidth]{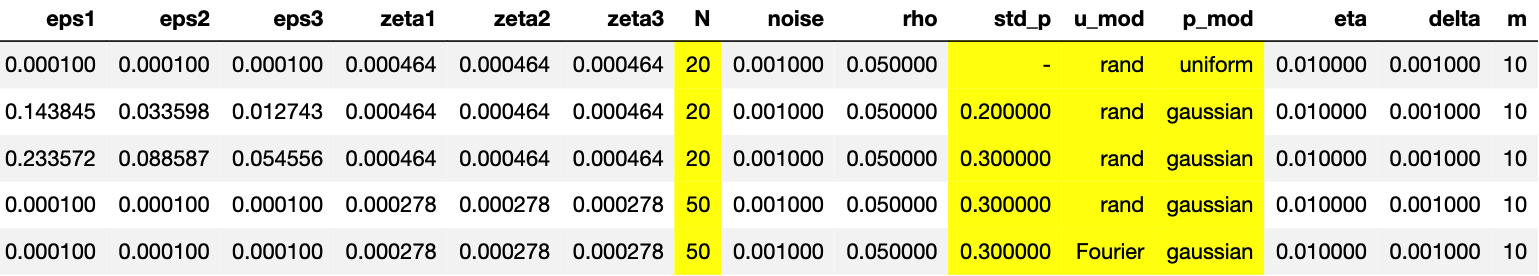}
\end{center}	
\caption{Checking observability with random input.} \label{figdfu}
\end{figure*}

\ \\ In the previous results, the control profile was systematically taken to be a randomly sampled truncated Fourier series. In order to evaluate whether this assumption is very specific regarding the certifiable reconstruction precisions, Figure \ref{figdfu} shows the results for uniformly randomly generated profiles inside the admissible set $[0.049, 0.449]$. It can be observed that while the same orders of magnitude are obtained, one can observe by comparing the first lines of Figures \ref{figdfp} and \ref{figdfu} that random profiles seems to enhance the observability at least for the setting that is common to these two lines. 

\ \\ 
\subsubsection{Impact of the certification precision parameter}
\begin{figure*}
\begin{center}
 \includegraphics[width=\textwidth]{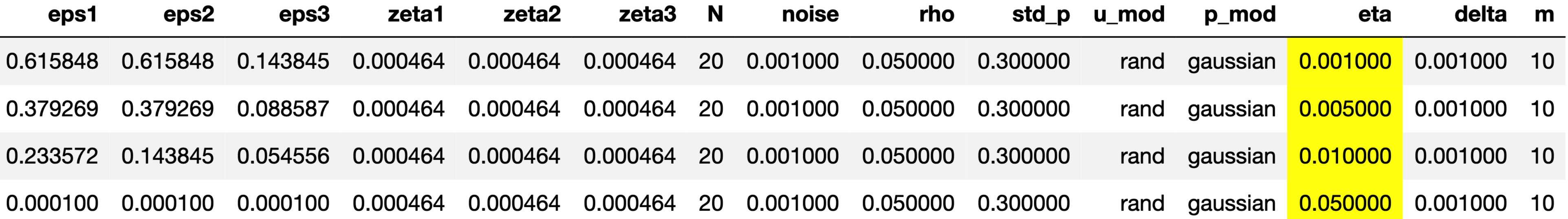}
\end{center}	
\caption{Impact of the precision $\eta$ required in the probabilistic certification.} \label{figdfeta}
\end{figure*}

\ \\ Figure \ref{figdfeta} shows how the results change when the targeted precision $\eta$ of the probabilistic certification is degraded. As expected, one can see that under the settings of this figure (the same for all values of $\eta$), a certification precision of $\eta=10^{-3}$ seems to lead to a certifiable upper bounds on the estimation errors on the observation-target variables which are quite high (roughly useless given the definition of the set $\mathbb X$). This is obviously due to the high level of parameter dispersion $s_{td}=0.3$. The certifiable  reconstruction error decreases when $\eta$ is increased meaning that the part of the pairs over which the reported reconstruction error are guaranteed is a smaller set of admissible pairs. For instance, the last line of Figure \ref{figdfeta} indicates that up to $5\%$ of the samples correspond to the presence of indistinguishable pairs. For the remaining $95\%$ of the cases, an almost zero reconstruction error can be certified (provided that the optimization problem is correctly solved). 
\\
\subsubsection{Computation times and scalability} \label{seccpu}
\begin{figure}
\begin{center}
\includegraphics[width=0.35\textwidth]{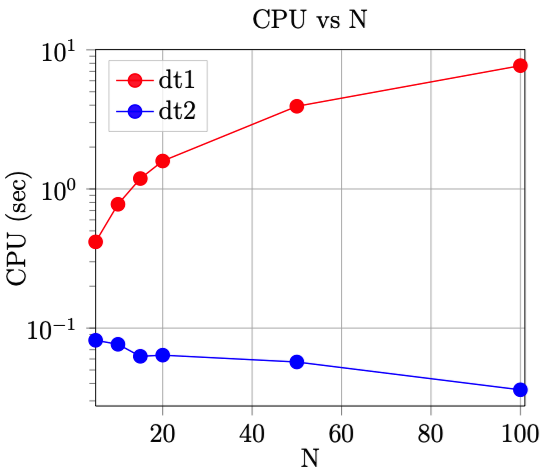}
\includegraphics[width=0.35\textwidth]{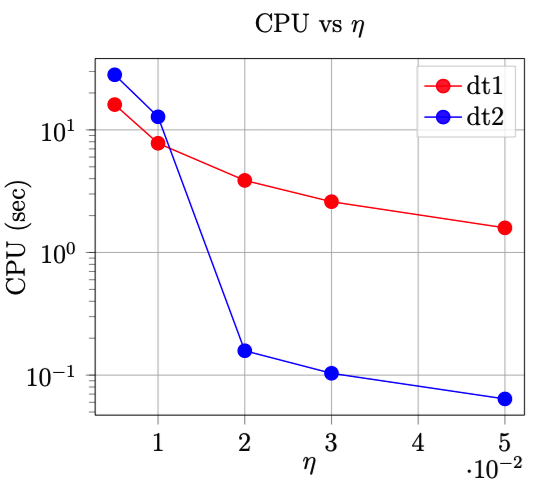}
\end{center}	
\caption{{\bf Computation times} using $\delta=0.001$, $m=10$. (top) cpu (sec) versus the observation horizon $N$ for $\eta=0.01$. (bottom) cpu (sec) versus $\eta$ for an observation horizon of $N=20$.} \label{figcpu}
\end{figure}

\noindent As it has been mentioned in Section \ref{seccomplexity}, the computation time is the sum of the cpu times needed for two successive tasks, namely:
\begin{itemize}
\item {\bf dt1}: the time needed to generate the scenarios 
\item {\bf  dt2}: the time needed to find the optimal design parameter. 
\end{itemize}
These times are mainly impacted by the following parameters:
\begin{enumerate}
\item The number of scenarios to be simulated which is defined by the four parameters $\eta$, $\delta$, $n_\Theta$ and $m$ through the expression (\ref{ineqN}) that determines the number of scenarios to be simulated. This linearly affects {\bf  dt1} but not {\bf  dt2}. \\
\item The time needed to simulate a single scenario which is for a given system depends on the observation horizon $N$ that impact the cpu time {\bf dt1}. This dependence is linear if a fixed step integration scheme is used to integrate ODE's models. \\
\item The search algorithm that is adopted to find the optimal design parameter. In the above computation, a simple alphabetic search is adopted with increasing $\epsilon=\theta_1$. 
\end{enumerate}
Note that the formulae (\ref{ineqN}) does not depend on the state or the parameter vector's dimension. This means that the size of the system and the number of its uncertain parameters do not lead to an exponential increase in the computation time, only the simulation time would affect linearly the computation time of the certification scheme. Note also that both $n_\Theta$ and $\delta$ appears logarithmically in the expression (\ref{ineqN}). This same expression shows clearly that increasing the confidence of the certification be reducing $\delta$ linearly increases the number of scenarios. \\ \ \\  
Regarding the scalability issue, it is worth underlying that the nature of the computation is probably the one that is most prone to parallelization as different scenarios are simulated independently in the first task while different values of $\theta$ are combined with each scenario to compute the constraint violation indicator. This might induce a high parallelization rate which together with the possibility to use efficient optimized compiled code (while python is used here) suggests that the scalability is far away from being an issue here. \\ \ \\ 
As far as the example is concerned, Figure \ref{figcpu} shows the computation times for the certification scheme as a function of the observation horizon or as a function of the certification precision $\eta$. The certification confidence parameter $\delta=0.001$ and the number of scenarios with failures $m=10$ are used. Note how the cpu time for the second task (optimizing the design parameter) decreases with the observation horizon as the number of values to be visited (in alphabetic order) is reduced because of the observability gained by the use of higher observation horizon. 

\section{Conclusion and future possible investigation} \label{secConc}
This paper proposes a general scalable scheme for the analysis of observability and parameter reconstruction in the context of nonlinear dynamical systems that are subjected to parametric uncertainties and measurement noise. The observability is taken in a more general sense than the standard (extended) observability commonly used in the sense that it is the possibility and the precision to which it is possible to reconstruct  specific expressions of the state/parameter that can be investigated by the proposed scheme. \\ \ \\  
In addition to the analysis power the proposed scheme offers, it can also be used as a tool to optimally design the parameters of the observation scheme and/or to specify the requirements in terms of the quality of the knowledge and the level of noise of the sensors being used to achieve a pre-specified level of reconstruction precision. It can also be used as a tools for expriment design, namely, what are the excitation profiles that lead to better identifiability of the model's parameters. \\ \ \\ 
It is mandatory to keep in mind that the certification results do not hold for a specific implementation of optimizers. The results of this paper concern the case where a perfect optimizer is available. In this sense, the scheme should be viewed as a way to answer the observability question as a property of the uncertain dynamic system with imperfect sensors and uncertain parameters. It is by no means a way to assess the degree of actual success in reconstructing the observation target variables under a specific solver that implements the MHE principle through a specific algorithm that comes itself with its own imperfections, choices and undesired behavior in the presence of local minima. \\ \ \\ 
A possible continuation of the present work concerns the investigation of the regions of the space of state and parameters where the  certification constraint does not hold leading to degraded certification results. This can be an important step in the analysis since these regions of the space might have been wrongly included while they are obviously to be excluded by the very definition of the operational space of the system. In such cases, these regions should be removed and the computation re-done in order to come out with more consistent results. \\ \ \\ Another undergoing work consists in applying the proposed scheme to standard models that are widely used in control and analysis of biological systems (diabetes, cancer, HIV, pandemics propagation models, etc) since these systems are by nature defined up to the knowledge of a high number of highly uncertain parameters. \\ \ \\ 
Finally, the program codes that served in producing the results of this paper will  shortly be available in the GitHub site of the author\footnote{https://github.com/mazen-alamir}.

\bibliographystyle{plain}
\bibliography{almost_obs.bib}

\end{document}